\documentclass[12pt]{article}

\usepackage{geometry}
\usepackage{graphicx}
\usepackage{subfig}
\usepackage{multirow}
\usepackage{caption}
\usepackage{floatrow}
\usepackage{float}
\usepackage{amsfonts}
\usepackage{amsmath}
\usepackage[%
    colorlinks=true,
    pdfborder={0 0 0},
    linkcolor=red]{hyperref}
\usepackage[english]{babel}
\usepackage{cite}
\usepackage{soul}
\usepackage[normalem]{ulem}

\usepackage{subcaption} 
\usepackage[export]{adjustbox} 
\usepackage{cleveref} 

\begin{document}
\title{Twist Angle Dependent Ultrafast Transient Dynamics of MoSe$_2$/WSe$_2$ van der Waals Heterostructures beyond the Exciton Mott Transition}
\author{Vikas Arora$^{1,2}$, Pramoda K Nayak$^{3,4}$, D. V. S. Muthu$^{1}$, A K Sood$^{*1,2}$}
\date{\today}
\maketitle

\begin{center}
\textit{$^{1}$Department of Physics, Indian Institute of Science, Bangalore 560012, India\\
$^{2}$Centre for Ultrafast Laser Applications, Indian Institute of Science, Bangalore 560012, India\\
$^{3}$2D Materials Research and Innovation Group, Department of Physics, Indian Institute of Technology Madras, Chennai 600036, India \\
$^{4}$Centre for Nano and Materials Sciences, Jain (Deemed-to-be University), Jain Global Campus, Kanakpura, Bangalore 562112, Karnataka, India}
\end{center}

\begin{center}
\end{center}

\begin{abstract}
Two-dimensional van der Waals heterostructures (HS) exhibit twist-angle ($\theta$) dependent inter-layer charge transfer, driven by moir{\'e} potential that tunes the electronic band structure with varying $\theta$. Apart from the magic angles of $\sim$3 $^\circ$ and $\sim$57.5 $^\circ$ that show flat valence bands (twisted WSe$_2$ bilayer), the commensurate angles of 21.8$^\circ$ and 38.2$^\circ$ reveal the Umklapp light coupling of interlayer excitons. We report our results on non-degenerate optical pump-optical probe spectroscopy of MoSe$_2$/WSe$_2$ HS at large twist angles under high photoexcitation densities above the Mott transition threshold, generating interlayer localized charge carriers. We show that the recombination time of electrons and holes is minimum at the commensurate angles. The strength of non-radiative interlayer Auger recombination also shows a minimum at the commensurate angles. The fluence dependence of interlayer carrier recombination time suggests additional relaxation channels near the commensurate angles. 
This study emphasizes the significance of the large twist angle of HS in developing transition metal dichalcogenides-based optoelectronic devices.

\end{abstract}

\section*{Introduction}
Two-dimensional materials such as graphene, transition metal dichalcogenides (TMDs), and h-BN, along with their heterostructures (HS), show many exciting physical properties. For example, in the case of TMDs, properties such as the large binding energy of excitons\cite{Alexey2014, Keliang2014,YuanyuanLi}, a strong non-linear optical response\cite{Nardeep2013,Yelei2013,Leandro2013} and spin-valley coupling\cite{Hualing2012,Kin2012} are exhibited. For the heterobilayers MoX$_2$/WX$_2$, the energy levels are decided by \textit{d} orbitals of transition metal. The \textit{5d} orbital of W is higher than that of the \textit{4d} orbital of Mo, and hence, the conduction band minima and valence band maxima of tungsten are higher than that of molybdenum. This creates a staggered or type-II band alignment\cite{Cheng2013,Pramoda2017}. This alignment facilitates ultrafast charge transfer between the layers, generation of interlayer exciton with longer decay times\cite{Torben2021,Schaibley2017}, and energy transfer among the layers\cite{Yan2022,YuanyuanLi}. The optoelectronic properties have been shown to depend on the twist angle between the layers. The twist between the two layers gives rise to a periodic array of potential minima known as the moiré pattern \cite{Tran_2019,Jin_2019,McHugh_2023,Niclas2022}. At low temperatures and low twist angles, the multiple sharp peaks superimposed over the photoluminescence (PL) spectra denotes the interlayer excitons trapped in moiré quantum wells\cite{Zidong2021, Michael2021}. 
Self-organized quantum dots are observed in MoSe$_2$/WSe$_2$ and MoS$_2$/WS$_2$ HS for twist angles less than 0.5$^\circ$ \cite{Enaldiev_2022}. For WS$_2$/MoSe$_2$ HS, twist angle tunes the moiré reciprocal lattice and hence the properties of moiré excitons such as oscillator strength and inter/intralayer mixing \cite{Zhang_2020}. In a twisted WSe$_2$ bilayer, flat top moiré valence bands are observed at magic angles of $\sim$3$^\circ$ and $\sim$57.5$^\circ$ \cite{Zhang2_2020,Devakul_2021,Naik_2018}. While small twist angles reveal the properties of moiré excitons, large twist angles describe the characteristics of interlayer excitons.
\par
Nayak \textit{et al.} \cite{Pramoda2017} observed the quenching of photoluminescence for the MoSe$_2$/WSe$_2$ HS for all the twist angles as compared to constituent monolayers. A redshift in PL energy of up to $\sim$100 meV has been observed in HS WSe$_2$/MoSe$_2$ depending on the twist angle \cite{Palekar_2024,Kai2016}. 
It has been reported that the PL intensity is two orders lesser for $\theta$ = 20$^\circ$ as compared to $\theta$ = 2$^\circ$ for MoSe$_2$/WSe$_2$ HS \cite{Kyle2019}. 
The interlayer distance and binding energy of interlayer exciton are twist-dependent and are maximum at 30$^\circ$, as revealed by the PL studies \cite{KeWu2021,Lishu2020}. Yu \textit{et al.} have shown that the interlayer excitons undergo Umklapp recombination near the commensurate twist angles of 21.8$^\circ$ and 38.2$^\circ$ \cite{Kyle2019, Hongyi2015}. The additional channel for recombination of interlayer excitons at these particular angles predicts anomalous relaxation dynamics for the HS. The significance of the commensurate angle 21.8$^\circ$ has been emphasized in recent reports on HS MoS$_2$/MoSe$_2$, demonstrating the highest thermoelectric performance (ZT = 2.96 at T = 700 K)\cite{Xiong_2024} and recording the highest indirect bandgap\cite{Sachin_2022}.
\par
A recent report on ultrafast electron diffraction of MoSe$_2$/WSe$_2$ HS indicates that interlayer heat transfer occurs on a $\sim$20 ps timescale, mediated by nonthermal phonons followed by interlayer charge transfer and scattering\cite{Johnson_2024}. The ultrafast transient absorption spectroscopy of TMDs HS has shown that the interlayer charge transfer happens on the order of 100 fs time scale\cite{Hong2014,Ceballos2014,Ceballos2015,Peng2016,Peymon2018,Krause2021}.  A slight variation in geometry or twist angle of the HS can vary the charge transfer time from 100 fs to 1 ps\cite{Ziheng2017, Jin2017,YuanyuanLi}. Zhu \textit{et al.} have reported the recombination time for interlayer exciton in twisted WSe$_2$/MoS$_2$ HS, varying from 40 ps to 3 ns for twist angles between 0$^\circ$ and 30$^\circ$, and attributed this variation to non-radiative defect-mediated recombination \cite{Zhu_2017}. In this study, no systematic twist-angle dependence of recombination time was observed. We note that these experiments were performed at a fluence of $\sim$10$^{11}$cm$^{-2}$, which is below the exciton Mott transition (MT) \cite{Zhu_2017}. The Mott transition occurs above a certain photoexcitation density, where the distance between excitons becomes smaller than their radius, thus screening the Coulomb interactions between electrons and holes resulting in an unbounded electron-hole plasma \cite{Hendry_2007, Kirsanske_2016}. The Mott transition of excitons has been observed in MoSe$_2$/WSe$_2$ HS, where interlayer excitons transform into an electron-hole plasma localized within the individual layers at a photoexcitation density of $\sim$4$\times$10$^{12}$ cm$^{-2}$,  \cite{Wang_2019}. An abrupt change in the interlayer exciton photoluminescence linewidth has been observed above the MT photoexcitation density \cite{Stern_2008,Wang_2019}. There have been many reports on different heterostructures for carrier relaxation dynamics but primarily for small twist angles \cite{Ziheng2017,Albert2015,Jin2017}, with only a few focusing on large twist angles \cite{Zhu_2017}. All these studies are focused on the photoexcitation densities below the MT. This has motivated us to explore the relaxation dynamics above the exciton Mott transition in TMDs HS with large twist angles.
\par
Here, we report the non-degenerate optical pump-optical probe studies of MoSe$_2$/WSe$_2$ HS at different twist angles at ambient conditions. A significant decrease in the recombination time of interlayer electron-hole plasma near the commensurate angles of 21.8$^\circ$ and 38.2$^\circ$ is observed, which is attributed to additional radiative recombination channels arising from Umklapp recombination. Further, the strength of the interlayer Auger recombination is minimum near the commensurate angles.
Our present study can help towards the application of type-II HSs in designing optoelectronic devices, namely photovoltaics, photodetectors, and light-emitting devices where charge transfer and recombination times are important parameters \cite{Liu2016}.

\section*{Experimental details}
Non-generate optical pump-optical probe measurements were performed using Ti: Sapphire ultrafast laser system (1 kHz repetition rate, $\sim$50 fs pulse width, M/s Spectra-Physics, Spitfire Ace) with the pump and probe energies of 3.1 eV and 1.55 eV, respectively. The pump and probe beams are passed collinearly through an objective lens with the NA = 0.9. The spot size of the beam at the sample point was measured to be $\sim$5 $\mu$m. The spot size is the same as the overlapping region of the heterostructure, and the resulting differential reflectivity ($\Delta R/R$) covers the entire overlapping region. Experiments were carried out for different twist angles of MoSe$_2$/WSe$_2$ HS at two pump fluences of 2.0 and 2.8 mJ/cm$^2$ at room temperature. To investigate carrier relaxation dynamics above the exciton Mott transition, we use a probe beam with a fixed wavelength of 800 nm.
\par
\begin{figure}[H]
  \includegraphics[width=\textwidth]{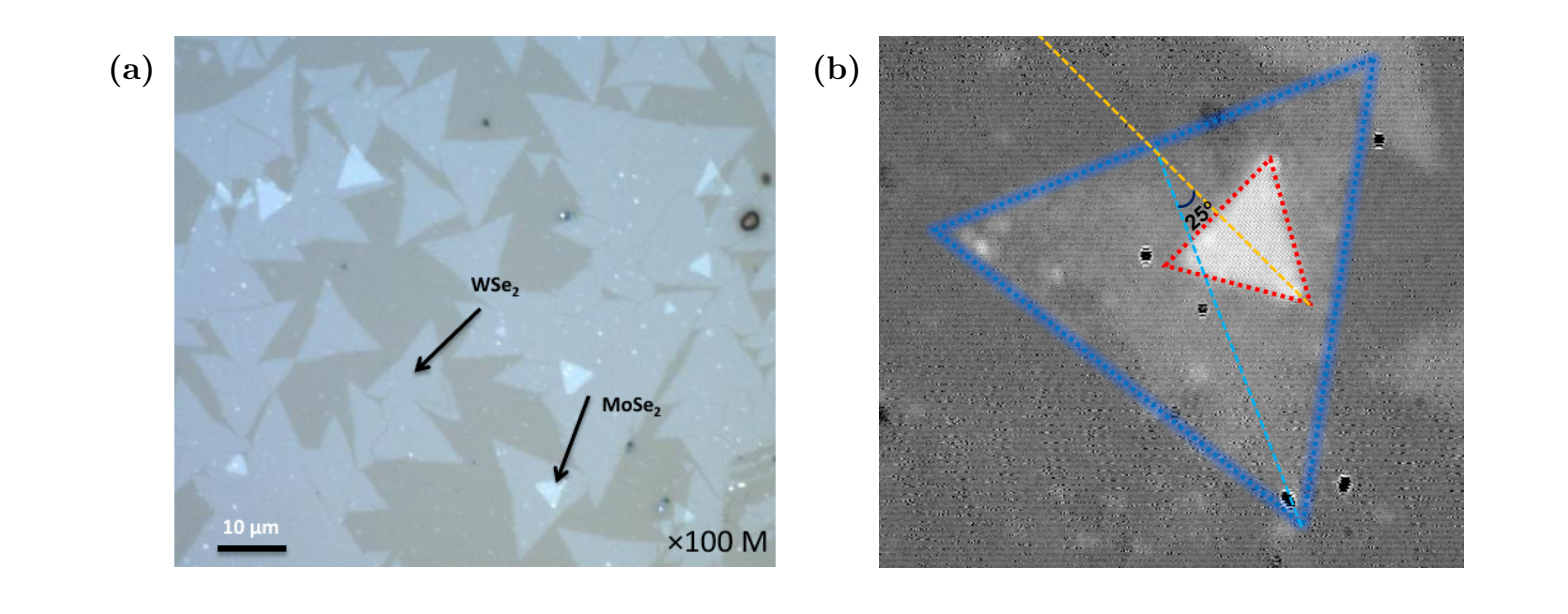}
  \caption[Microscopic image of MoSe$_2$/WSe$_2$ heterostructure]{(a) Microscopic image of MoSe$_2$/WSe$_2$ HS depicting WSe$_2$ as a bottom layer (big triangles) with dull contrast and MoSe$_2$ as a top layer (smaller triangles) with brighter contrast, (b) Angle evaluation between two monolayers with the blue(red) triangle as a boundary of WSe$_2$(MoSe$_2$) monolayers.} 
\label{Fig:6p1}
\end{figure}

 
The MoSe$_2$/WSe$_2$ HS were prepared using the chemical vapor deposition method on a \textit{c}-plane sapphire substrate via a two-step process similar to our earlier work \cite{Pramoda2017}. The sapphire substrate were chosen because the strain effects are expected to be minimal (0.08$\%$) compared to silica substrate (0.38$\%$) \cite{Hang_2022}. The individual monolayers of MoSe$_2$ and WSe$_2$ have a triangular geometry as shown in the microscopic image of the sample (Figure \ref{Fig:6p1}\textcolor{red}{a}). The triangles with light (bright) contrast and bigger (smaller) in size represent WSe$_2$ (MoSe$_2$) monolayers as depicted by the black arrows in Figure \ref{Fig:6p1}\textcolor{red}{a}. A random alignment of triangular-shaped monolayers was obtained as a result of the CVD process, providing us with various twist angles. It has been verified that the TMDs with triangular geometry terminate with transition metal atoms at the edges using transmission electron microscopy\cite{vander2013} and scanning tunneling microscopy\cite{Lauritsen2007}. Therefore, the relative orientation of the layers provides us with the angle between the two layers as there is a direct correlation of crystal orientation with the triangular geometry of each monolayer\cite{Ziheng2017, vander2013, Zhou2015}. The twist angle is evaluated as the angle between two lines collinear to the centroid and the vertex of the triangle\cite{LiuK2014}. One example of evaluating the twist angle is presented in Figure \ref{Fig:6p1}\textcolor{red}{b}, where red and blue outlined triangles represent MoSe$_2$ and WSe$_2$ monolayers, respectively. The lines collinear to the vertex and centroid are the angle bisector in the case of equilateral triangles, which gives the twist angle. The schemes for the front and side view of MoSe$_2$/WSe$_2$ HS for different twist angles are shown elsewhere \cite{Pramoda2017}. 

\section*{Raman characterization}
\begin{figure}[H]
    \includegraphics[width=\textwidth]{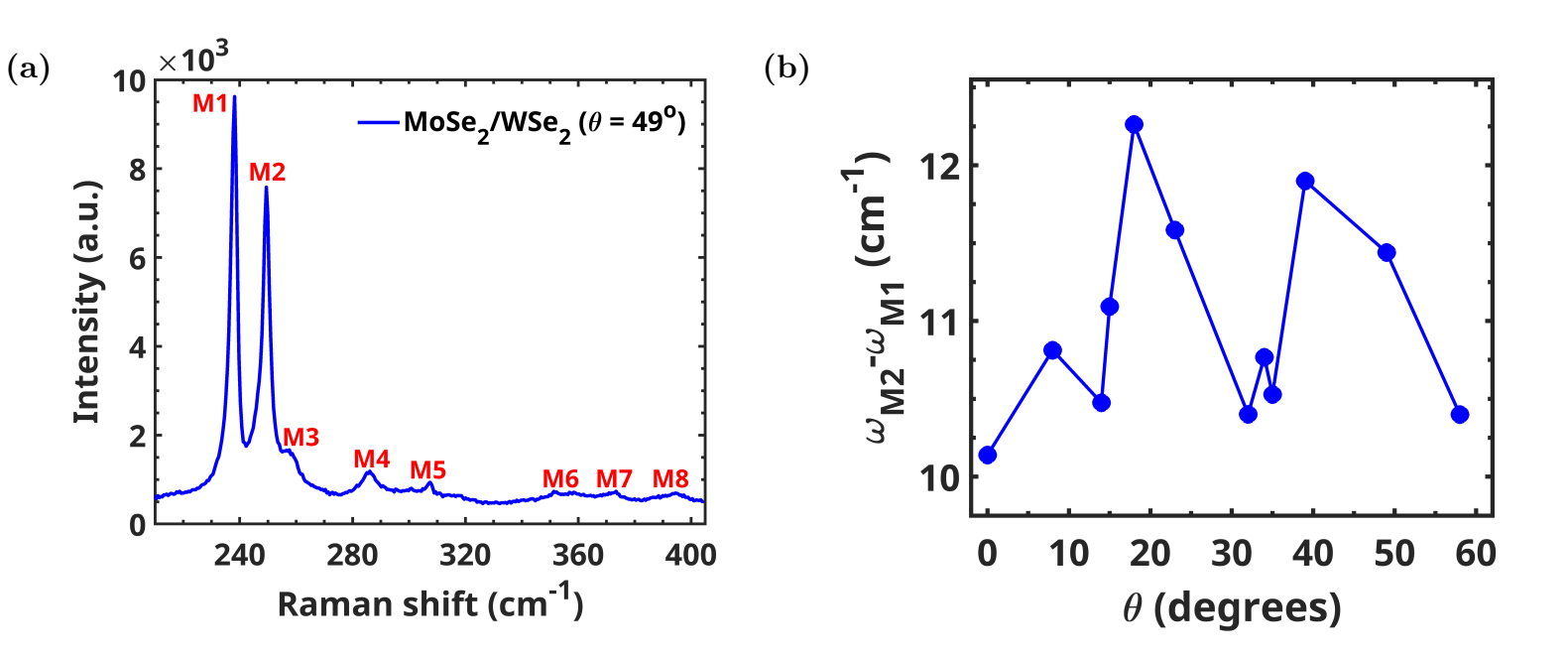}
    \caption[Raman characterization of HS and analysis of M1 and M2 Raman modes with different twist angles]{(a) The Raman spectrum for HS at a $\theta$ = 49$^\circ$. M1(M2) represents the A$_{1g}$ mode of MoSe$_2$(WSe$_2$), The respective modes are as follows:  M1(238.5 cm$^{-1}$), M2(249 cm$^{-1}$), M3(258.5 cm$^{-1}$), M4(286 cm$^{-1}$), M5(305 cm$^{-1}$), M6(353 cm$^{-1}$), M7(373 cm$^{-1}$), M8(394 cm$^{-1}$). (b) The difference of frequencies of modes M1 and M2 is presented with twist angles. The difference is maximum in the proximity of commensurate angles of 21.8$^\circ$ and 38.2$^\circ$. The error bars are smaller than the symbols representing the data points.}
    \label{Fig:6p2}
\end{figure}

The blue and red shifts of the A$_{1g}$ modes in WS$_2$ and MoS$_2$ within their twisted HSs \cite{Lishu2020} motivated us to perform Raman spectroscopy on HS MoSe$_2$/WSe$_2$ for various twist angles. The group theory for the individual layer provides the irreducible decomposition of Raman active vibration modes A$_{1g}$(R), E$_{2g}$(R), and E$_{1g}$ (IR+R) for MX$_2$ monolayer \cite{YangM2017}. Figure \ref{Fig:6p2}\textcolor{red}{a} shows the Raman spectrum of the HS region for a twist angle of 49$^\circ$ recorded using 532 nm laser wavelength. We obtain the following vibration modes as shown in Figure \ref{Fig:6p2}\textcolor{red}{a}: M1(MoSe$_2$, A$_{1g}$): 238.5 cm$^{-1}$, M2(WSe$_2$, A$_{1g}$+E$_{2g}$): 249 cm$^{-1}$, M3(WSe$_2$, 2LA): 258.5 cm$^{-1}$, M4(MoSe$_2$, E$_{2g}$): 286 cm$^{-1}$, M5(WSe$_2$, A$_2''$): 305 cm$^{-1}$, M6(MoSe$_2$, A$_2''$): 353 cm$^{-1}$, M7: 373 cm$^{-1}$ and M8: 394 cm$^{-1}$ \cite{Woosuk2019, Lama2021}. It confirms the formation of HS with the presence of all the modes of individual monolayers of MoSe$_2$ and WSe$_2$. From the perspective of the charge transfer among two vertically stacked monolayers, we concentrate on the out-of-plane modes (A$_{1g}$) of the two monolayers MoSe$_2$ and WSe$_2$ stacked together, namely M1 (MoSe$_2$) and M2 (WSe$_2$).
\par
Figure S1 (supplementary information) shows the Raman spectra of M1, M2, and M3 modes with normalized intensities for different twist angles. The black dashed lines represent the shift of the peak position of modes M1 or M2 at different twist angles. We observed the softening of M1 and hardening of M2 mode for any non-zero twist angle as compared to the aligned geometry. Mode M2 starts stiffening as the twist angle is varied till 17$^\circ$, followed by softening till 32$^\circ$ and then stiffening again till 39$^\circ$ which is again followed by softening. In a counter behavior, the M1 mode softens as the twist angle is varied till 17$^\circ$, followed by stiffening till 32$^\circ$ and then softening till 39$^\circ$, which is again followed by stiffening. The softening and stiffening of the A$_{1g}$ mode (both M1 and M2) depend upon the carrier doping as well as the interlayer coupling \cite{LiuK2014, Zhang2015, Chakraborty2012}. In transition metal dichalcogenide MoS$_2$, increasing the number of layers stiffens the A$_{1g}$ mode and softens the E$_{2g}$ mode due to increased dielectric screening of Coulomb interactions \cite{Lee_2010, Sanchez_2011}. It has been reported that the unidirectional transfer of electrons from monolayer WS$_2$ to monolayer MoS$_2$ leads to higher carrier density in MoS$_2$ and lower carrier density in WS$_2$, which can soften or stiffen the respective A$_{1g}$ modes\cite{Lishu2020}. For the similar staggered bands configuration of HS MoSe$_2$/WSe$_2$, unidirectional electron flow from WSe$_2$ to MoSe$_2$ increases (reduces) the carrier density in MoSe$_2$ (WSe$_2$) and leads to the softening (stiffening) of M1 (M2) modes. To understand the importance of twist angle, we have plotted the difference between the peak position of modes M1 and M2 with twist angles as shown in Figure \ref{Fig:6p2}\textcolor{red}{b}. Here, the solid blue line shows a systematic variation of the shift of M1 with respect to M2 (or vice-versa) with maximum shift near the commensurate angles of 21.8$^\circ$ and 38.2$^\circ$. For perfectly aligned HS ($\theta$=0$^\circ$), the electron transfer and the interlayer coupling lead to the softening of A$_{1g}$ of both M1 and M2 modes, if we compare these modes in monolayers of MoSe$_2$ and WSe$_2$ (Figure S1). However, with a non-zero twist angle, variation in the charge transfer and the interlayer coupling tunes the softening and the corresponding stiffening of M1 and M2 modes and shows an anomalous maximum shift near the commensurate angles. This can be easily understood as the maximum electron transfer and more dielectric screening of Coulomb interaction between electron and hole for the MoSe$_2$/WSe$_2$ HS at these specific angles, which are in the vicinity of commensurate angles. We also note the disappearance of M3 (2LA mode of WSe$_2$) at the twist angles of 19$^\circ$ and 39$^\circ$ as depicted by small arrows in Figure S1 (panel iii), which needs further investigation.

\section*{Results and discussions}

\begin{figure}[H]
    \includegraphics[width=\textwidth]{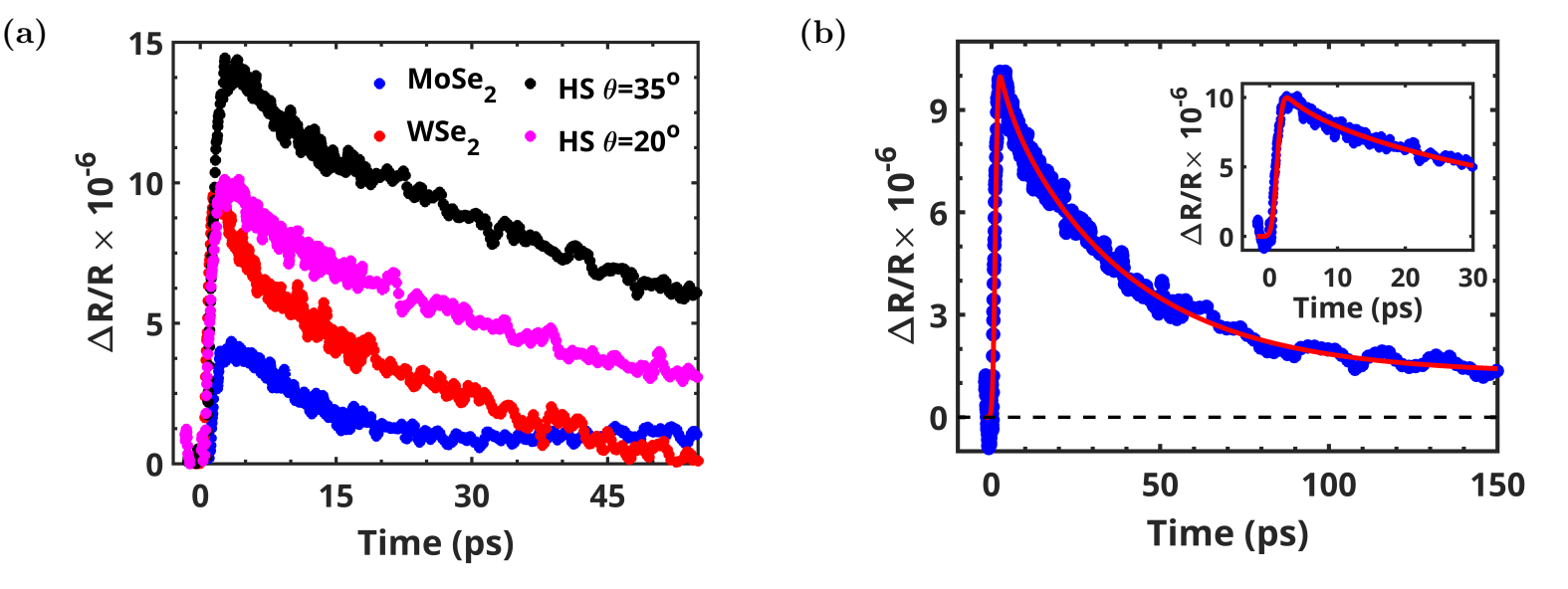}
    \caption[Differential reflectivity ($\Delta$R/R) of MoSe$_2$/WSe$_2$ HS at different twist angles]{(a) Time evolution of $\Delta$R/R for ML MoSe$_2$ (blue circles), ML WSe$_2$ (red circles), HS MoSe$_2$/WSe$_2$ at $\theta$=20$^\circ$ (pink circles), $\theta$=35$^\circ$ (black circles). (b) The differential reflectivity $\Delta$R/R for HS MoSe$_2$/WSe$_2$ at twist angle of $\theta$=20$^\circ$ (blue circles) is fitted using Eq.\eqref{eq:6p1} (red solid line). The inset shows the fit of the experimental data up to 30 ps.}
    \label{Fig:6p3}
\end{figure}

Figure \ref{Fig:6p3}\textcolor{red}{a} shows time-resolved differential reflectivity ($\Delta$R/R) for monolayer (ML) MoSe$_2$ (blue circles), ML WSe$_2$ (red circles), and HS MoSe$_2$/WSe$_2$ at twist angles ($\theta$) of 20$^\circ$ (pink circles) and 35$^\circ$ (black circles) with a fluence of 2.0 mJ/cm$^2$. It demonstrates that the relaxation dynamics for the HS differ from those of the individual MLs and vary with different twist angles. For ML MoSe$_2$, a single exponential decay provides a recombination time of 9.6$\pm$0.3 ps, whereas ML WSe$_2$ exhibits a biexponential decay with recombination times of 1.4$\pm$0.15 ps and 20.8$\pm$0.5 ps. A biexponential fit for the $\Delta$R/R(t) (blue circles) for the HS at $\theta$=20$^\circ$ is shown by solid red curve in Figure \ref{Fig:6p3}\textcolor{red}{b}, using the equation:
\begin{equation}
    \centering
    \frac{\Delta R}{R}(t)=\frac{1}{2}\left(1+erf\left(\frac{t-t_0}{\tau_{r}}\right)\right)(A_1e^{-t/\tau_1}+A_2e^{-t/\tau_2}+A_3)
    \label{eq:6p1}
\end{equation}
where $(1+erf((t-t_0)/\tau_r))$ represents the rise part of the time-resolved $\Delta$R/R, with $\tau_r$ being the rise time, which varies in the range of $\sim$0.7-1.0 ps for different twist angles. The fast and slow recombination processes are characterized by amplitudes and recombination times (A$_1$, $\tau_1$) and (A$_2$, $\tau_2$), respectively. A$_3$ is the strength of the slow relaxation process, taken to be constant in the delay time up to 150 ps. Before extracting the fitting parameters for various twist angles, we discuss the mechanism involved in the relaxation dynamics of HS MoSe$_2$/WSe$_2$.
\par
\begin{figure}[H]
   \centering
    \includegraphics[width=\textwidth]{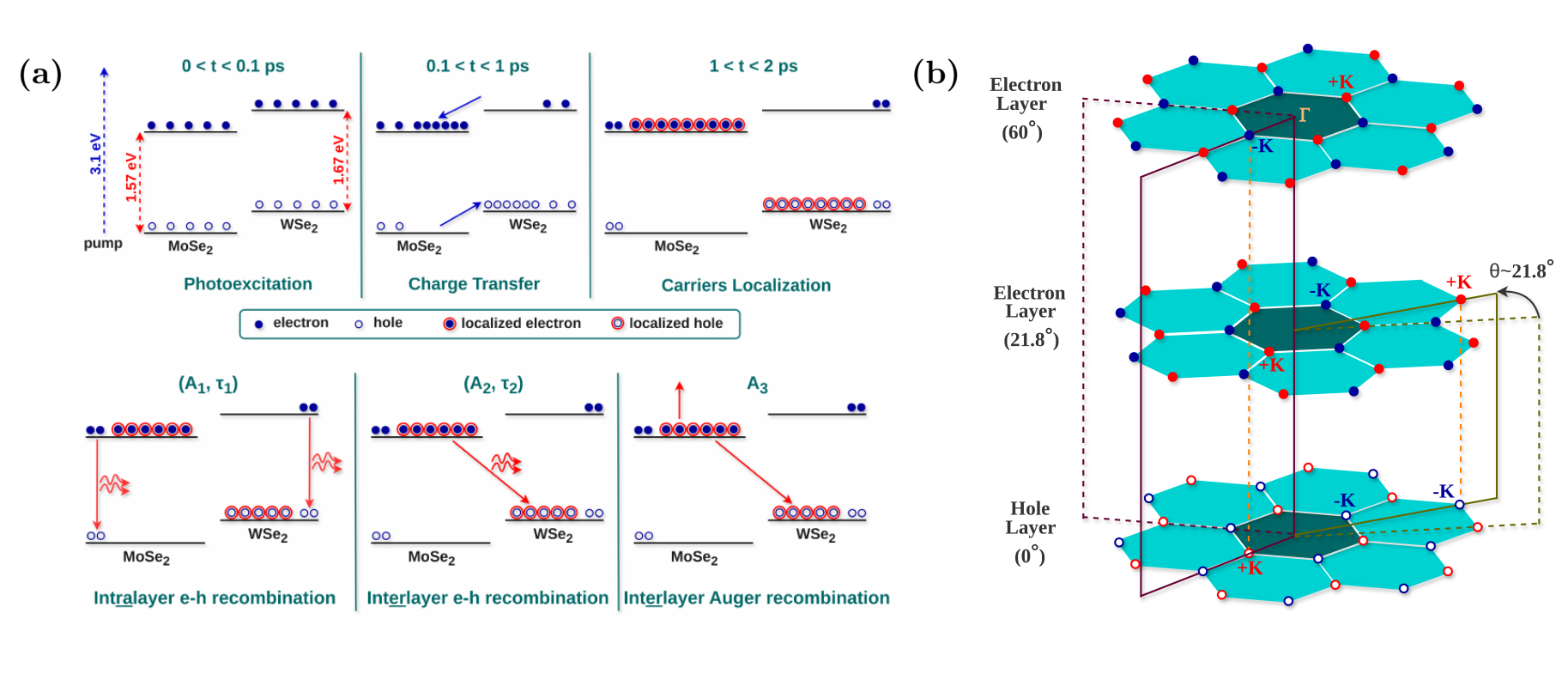}
    \caption[Ultrafast charge transfer in the MoSe$_2$/WSe$_2$ HS]{Type II band structure of HS MoSe$_2$/WSe$_2$, (a) Top panel: Photoexcitation: Electrons are transferred to the conduction band of MoSe$_2$ and WSe$_2$, Charge transfer: electrons (holes) transfer from conduction (valence) band of WSe$_2$ (MoSe$_2$) to MoSe$_2$ (WSe$_2$), Carriers Localization: interlayer charge carriers localized to distinct individual layers. Bottom panel: intralayer e-h recombination (A$_1$, $\tau_1$): electrons and holes within the same layer recombine radiatively, Interlayer e-h recombination (A$_2$, $\tau_2$): Electron and holes in different layers recombine radiatively, Interlayer Auger recombination (A$_3$): Electrons and holes in distinct layers recombine non-radiatively, transferring energy to a third particle, which is either an electron or a hole. (b) A schematic representation for the valley alignments in the extended Brillouin zone for a twisted HS: The solid and open dots represent the -K(blue) and +K(red) valleys in the electron and hole layers. For a twist angle of $\sim$21.8$^\circ$, the valleys +K, -K and -K, +K align in the second Brillouin zone (indicated by an orange dashed line from the middle layer to the bottom layer), which has the same valley pairing for the 60$^\circ$ twisted HS (denoted by the orange dashed line from the top layer to the bottom layers). Adapted from \textit{Seyler et al.} \cite{Kyle2019}.}
    \label{Fig:6p4}
\end{figure}

Figure \ref{Fig:6p4}\textcolor{red}{a} shows the type II band alignment in MoSe$_2$/WSe$_2$ HS depicting higher energy states of valence band maxima (VBM) and conduction band minima (CBM) of WSe$_2$ with respect to that of MoSe$_2$. The optical pump excitation using 400 nm (3.1 eV) laser pulse injects carriers into the conduction band much above the band gap for both constituent monolayers (MoSe$_2$ E$_g$=1.57 eV, WSe$_2$ E$_g$=1.67 eV) \cite{Lu_2014,Huang_2014}, as shown in the top left panel of Figure \ref{Fig:6p4}\textcolor{red}{a}. The energetic hot carriers undergo rapid electron-electron scattering and relax to the band edge, followed by a charge transfer between the MLs, creating both inter and intralayer excitons \cite{Hong2014,Ceballos2014}. However, at a photoexcitation density of 2 mJ/cm$^2$ ($n_0\sim$ 6.1$\times$10$^{14}$ cm$^{-2}$), these interlayer excitons undergo Mott transition, where they transform into a hot electron and hot hole plasma in MoSe$_2$ and WSe$_2$ monolayers, respectively \cite{Wang_2019,Stern_2008,Kirsanske_2016,Hendry_2007}. These carriers, localized to individual layers, emerge $\sim$1 ps after the photoexcitation, when most of the electrons (holes) get transferred to CBM (VBM) of MoSe$_2$ (WSe$_2$)  \cite{Wang_2019,Ceballos2014}, as shown in the top right panel of Figure \ref{Fig:6p4}\textcolor{red}{a}. Localized interlayer charge carriers take longer to recombine compared to intralayer carriers, as the electrons and holes are confined to different layers rather than being in the same monolayer. The bottom left panel of Figure \ref{Fig:6p4}\textcolor{red}{a} represents the radiative recombination of the intralayer electrons and holes in the individual monolayers, associated with (A$_1$, $\tau_1$). For WSe$_2$/MoSe$_2$ HS with non-zero twist angle (4$^\circ\pm$2$^\circ$), observation of interlayer PL for excitation densities above Mott transition \cite{Wang_2019, Palekar_2024} demonstrate the radiative recombination of interlayer electrons and holes, but only at a temperature of 4 K. However, Seyler \textit{et al.} have reported that interlayer electrons and holes in MoSe$_2$/WSe$_2$ do not align at $\pm$K valleys in the first Brillouin zone, emphasizing non-radiative recombination as a dominant process \cite{Kyle2019}. Therefore, we attribute the slow component of the relaxation dynamics (A$_2$, $\tau_2$) to indirect radiative (phonon-mediated) and non-radiative recombination of interlayer electrons and holes, as illustrated in the lower middle panel of Figure \ref{Fig:6p4}\textcolor{red}{a}. Motivated by the suggestion of non-radiative Auger recombination of interlayer excitons at high excitation densities \cite{Binder_2019, Cai_2024}, the A$_3$ component is attributed to non-radiative interlayer Auger recombination. In this process, electrons and holes from different MLs recombine and transfer the energy to the third particle, such as an electron or a hole, as demonstrated in the lower right panel of Figure \ref{Fig:6p4}\textcolor{red}{a}.

\par
\begin{figure}[H]
    \includegraphics[width=\textwidth]{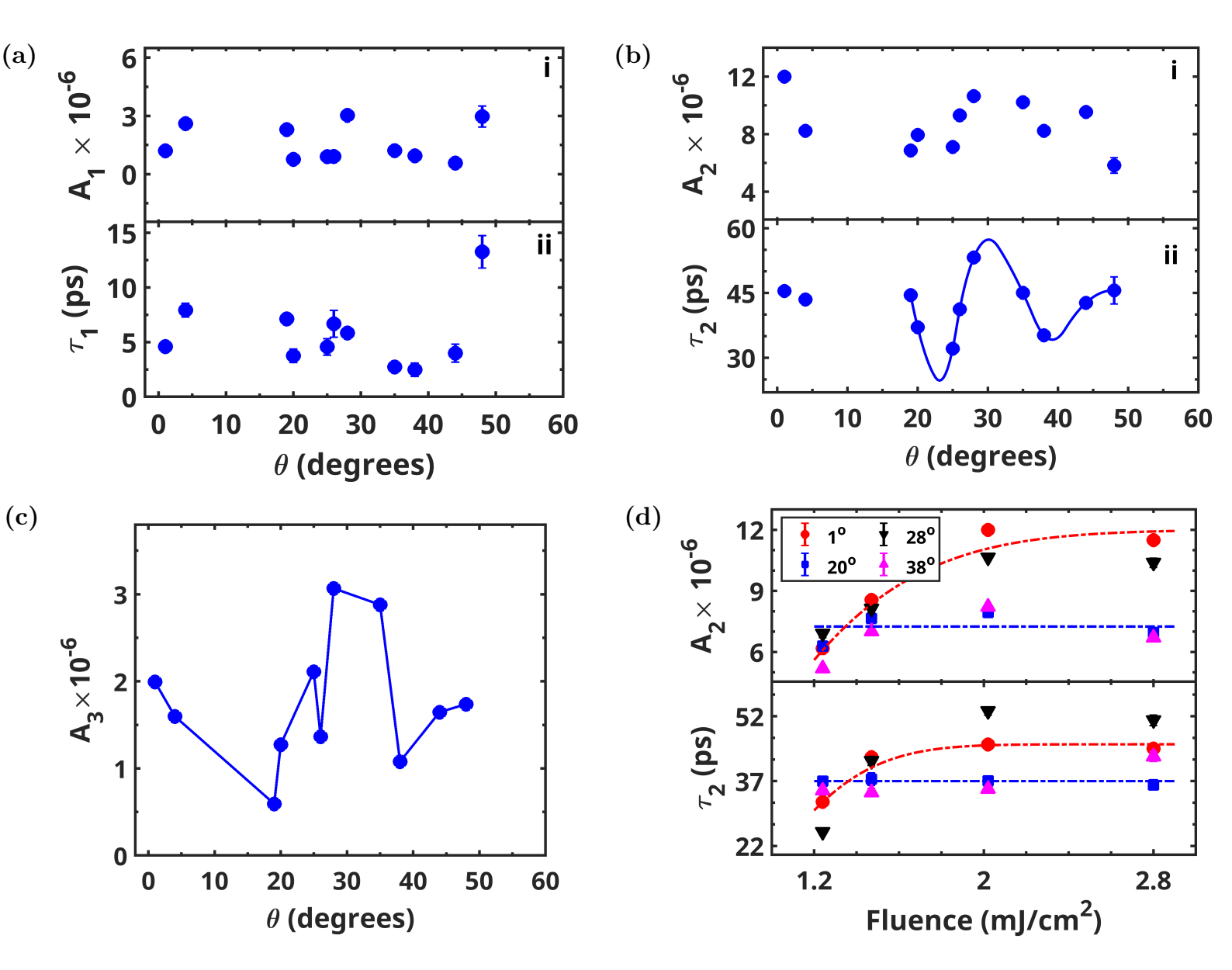}
    \caption[Relaxation dynamics parameters with twist angles]{Twist-angle dependence of (a) amplitude (A$_1$) and recombination time ($\tau_1$), (b) amplitude (A$_2$) and recombination time ($\tau_2$), the solid blue line represents a cubic spline. (c) amplitude (A$_3$) at a fluence of 2.0 mJ/cm$^2$.  (d) Fluence dependence of amplitude (A$_2$) and recombination time ($\tau_2$) for different twist angles. The dashed and dotted (red and blue) lines are guides to the eyes for the data corresponding to 1$^\circ$ and 20$^\circ$. The error bars are smaller than the symbols representing the data points.}
    \label{Fig:6p5}
\end{figure}

Figure \ref{Fig:6p5}\textcolor{red}{a} shows the twist-angle ($\theta$)-dependent behavior of the intralayer photoexcited carriers (A$_1$) and its recombination time ($\tau_1$) at a fluence of 2.0 mJ/cm$^2$. A$_1$ and $\tau_1$ exhibit no systematic dependence on twist angle as is expected for intralayer carriers recombination in individual monolayers. The recombination dynamics of the interlayer carriers is presented in Figure \ref{Fig:6p5}\textcolor{red}{b}, where amplitude A$_2$ is twist-angle independent. The recombination time of interlayer localized electrons and holes, $\tau_2$ (panel ii of Figure \ref{Fig:6p5}\textcolor{red}{b}) exhibits minima near 20$^\circ$ and 38$^\circ$. The blue dots represent the data, while the solid blue line is a cubic spline. The component associated with non-radiative Auger recombination of interlayer carriers, A$_3$, also shows minima for $\sim$20$^\circ$ and $\sim$38$^\circ$ (represented by the solid blue line in Figure \ref{Fig:6p5}\textcolor{red}{c}). The trend of the fitting parameters at a fluence of 2.0 mJ/cm$^2$ is similar at a fluence of 2.8 mJ/cm$^2$, as shown in supplementary Figure S2.
\par
We offer a qualitative understanding of our results. We observed that $\tau_2$ is maximum at the twist angle of $\sim$30$^\circ$ (Figure \ref{Fig:6p5}\textcolor{red}{b}), \textit{i.e.} the interlayer localized carriers take the longest time to recombine. This is due to the interlayer distance being maximum at the twist angle of 30$^\circ$ in TMDs HSs\cite{KeWu2021,Lishu2020}. Equally interesting is the observation of two minima of $\tau_2$ observed near 20$^\circ$ and 38$^\circ$. These twist angles are close to the commensurate angles of $\theta$ = 21.8$^\circ$ and 38.2$^\circ$ for MoSe$_2$/WSe$_2$ HS, where interlayer localized charge carriers or/and excitons can access an additional channel to recombine\cite{Kyle2019, Hongyi2015}. \textit{Seyler et al.} have demonstrated that for any non-zero twist angle between two layers, the holes in the valence band of WSe$_2$ and electrons in the conduction band of MoSe$_2$ at $\pm$K points do not align in the first Brillouin zone. However, at the commensurate angle of 21.8$^\circ$, the valence band and conduction band align at $\pm$K points in the second Brillouin zone, as shown in Figure \ref{Fig:6p4}\textcolor{red}{b}, adapted from ref. \cite{Kyle2019}. At these specific angles, the interlayer charge carriers can recombine radiatively, and the momentum mismatch (due to non-zero twist angle) is compensated by the reciprocal lattice of two layers; the process known as Umklapp recombination \cite{Hongyi2015, Kyle2019}. We attribute this Umklapp recombination in the vicinity of commensurate angles as a cause for the two minima observed for $\tau_2$ (Figure \ref{Fig:6p5}\textcolor{red}{b}). The two different fluences support the evidence of faster relaxation dynamics in the vicinity of commensurate twist angles. The additional radiative recombination channels for interlayer charge carriers result in lesser interlayer charge carriers involved in non-radiative Auger recombination, leading to a trend that shows minima in A$_3$ near these commensurate angles (Figure \ref{Fig:6p5}\textcolor{red}{c}). To further investigate the minima for $\tau_2$ near the commensurate angles, the relaxation dynamics is studied as a function of excitation density for different twist angles.

The fluence dependence of the relaxation dynamics of interlayer charge carriers (Figure \ref{Fig:6p5}\textcolor{red}{d}) shows an increase in interlayer charge carriers with increasing fluence, which saturates beyond 2.0 mJ/cm$^2$ indicating the threshold of charge transfer in HS MoSe$_2$/WSe$_2$ at fluences above 2.0 mJ/cm$^2$. It is important to note that $\tau_2$ is least fluence dependent at 20$^\circ$ and 38$^\circ$ in comparison to other twist angles, despite a significant increase in the number of interlayer charge carriers. Intralayer charge carriers do not exhibit this distinct behavior near commensurate angles, as shown by the fluence dependence of amplitude A$_1$ and $\tau_1$ in supplementary Figure S3. In conclusion, the additional recombination channels for interlayer charge carriers at commensurate angles demonstrate the fluence independence of recombination time ($\tau_2$) within the reported range up to 2.8 mJ/cm$^2$.


\section*{Conclusions}
In summary, we have reported the optical pump optical probe spectroscopy of MoSe$_2$/WSe$_2$ HS with different twist angles at high excitation densities above the Mott transition. We observed a significant decrease in interlayer exciton recombination time in the proximity of commensurate angles of 21.8$^\circ$ and 38.2$^\circ$. We interpret the result as a consequence of the availability of new radiative recombination channels due to Umklapp recombination. The least fluence dependence of interlayer charge carriers recombination time near commensurate angles further supports the presence of additional recombination channels. 
Our results can be helpful in designing the applications of twisted vdW HS in optoelectronic applications. 


\section*{Acknowledgments} 
AKS thanks the Department of Science and Technology, Government of India, for its financial support under the National Science Chair Professorship and Nanomission (Grant No. DST/NM/TUE/QM-5/2019). DVSM also acknowledges the DST for the Nanomission grant. VA acknowledges CSIR for the research fellowship. PKN acknowledges the MHRD STARS research grant [STARS/APR2019/396] and the support from the Institute of Eminence scheme at IIT-Madras through the 2D Materials Research and Innovation Group.


\floatsetup[figure]{style=plain,subcapbesideposition=top}

\clearpage

\clearpage

\clearpage

\newpage
\bibliographystyle{unsrt}
\bibliography{citations.bib}
\end{document}


\title{Twist Angle Dependent Ultrafast Transient Dynamics of MoSe$_2$/WSe$_2$ van der Waals Heterostructures beyond the Exciton Mott Transition}
\author{Vikas Arora$^{1,2}$, Pramoda K Nayak$^{3,4}$, D. V. S. Muthu$^{1}$, A K Sood$^{*1,2}$}
\date{\today}
\maketitle

\begin{center}
\textit{$^{1}$Department of Physics, Indian Institute of Science, Bangalore 560012, India\\
$^{2}$Centre for Ultrafast Laser Applications, Indian Institute of Science, Bangalore 560012, India\\
$^{3}$2D Materials Research and Innovation Group, Department of Physics, Indian Institute of Technology Madras, Chennai 600036, India \\
$^{4}$Centre for Nano and Materials Sciences, Jain (Deemed-to-be University), Jain Global Campus, Kanakpura, Bangalore 562112, Karnataka, India}
\end{center}

\begin{figure}[H]
    \includegraphics[width=\textwidth]{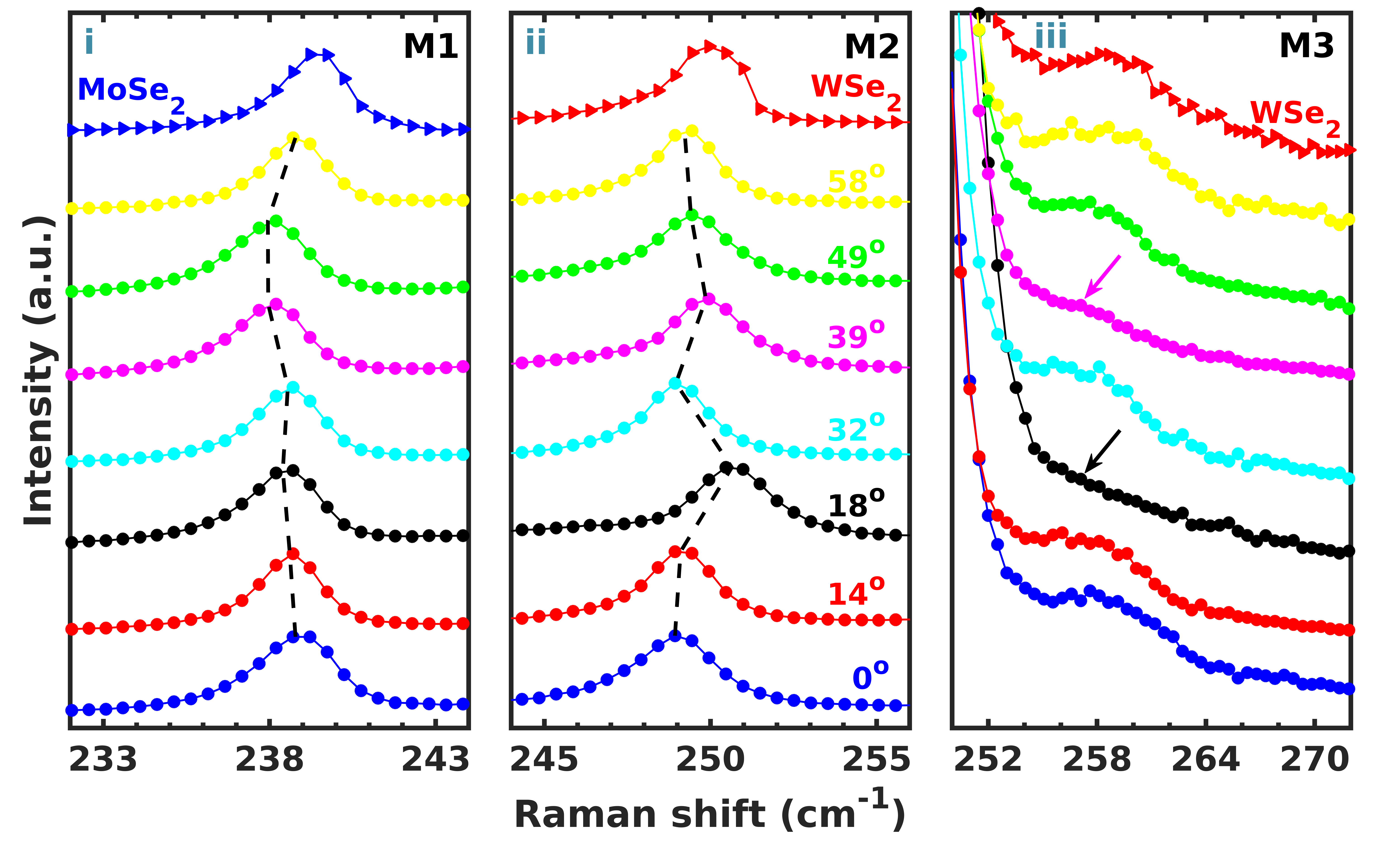}
    \renewcommand{\thefigure}{S\arabic{figure}}
    \caption[Raman spectroscopy of HS MoSe$_2$/WSe$_2$ for different twist angles]{The graphs depict (i) M1, (ii) M2 and (iii) M3 modes at different twist angles and constituent monolayers. The corresponding softening and stiffening of M1 and M2 modes with the twist angle are indicated by the black dashed line, M3 mode disappears at $\theta$=18$^\circ$ and 39$^\circ$ as indicated by solid arrows in panel iii. }
    \label{Fig:6pS1}
\end{figure}

\begin{figure}[H]
  \includegraphics[width=\textwidth]{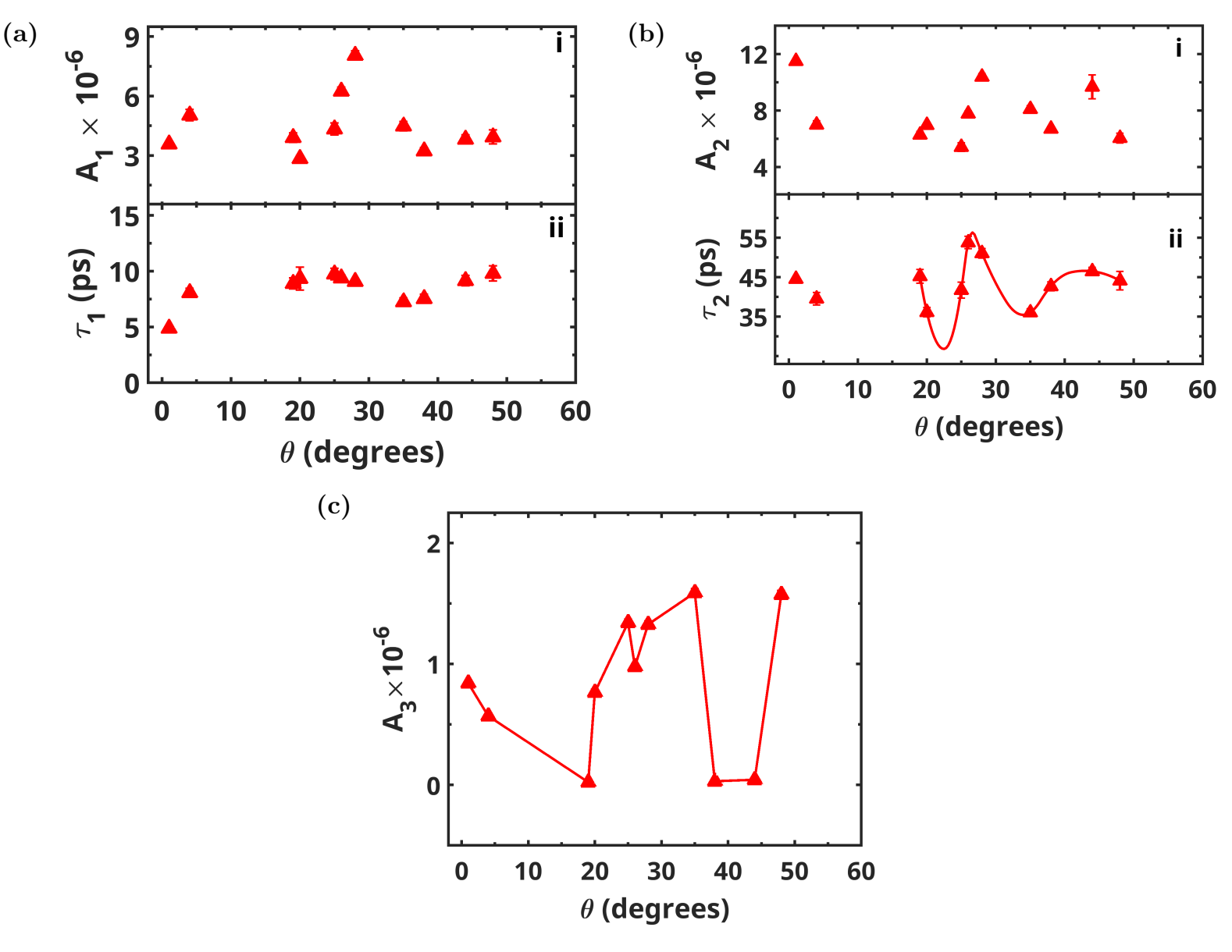}
    \renewcommand{\thefigure}{S\arabic{figure}}
  \caption[Amplitude of recombination time as a function of twist-angle]{Amplitude and recombination time: (a) (A$_1$, $\tau_1$), (b) (A$_2$, $\tau_2$), (c) amplitude (A$_3$) as a function of twist angle at a fluence of 2.8 mJ/cm$^2$.} 
\label{Fig:6pS2}
\end{figure}

\clearpage

\begin{figure}[H]
    \includegraphics[width=3.25in]{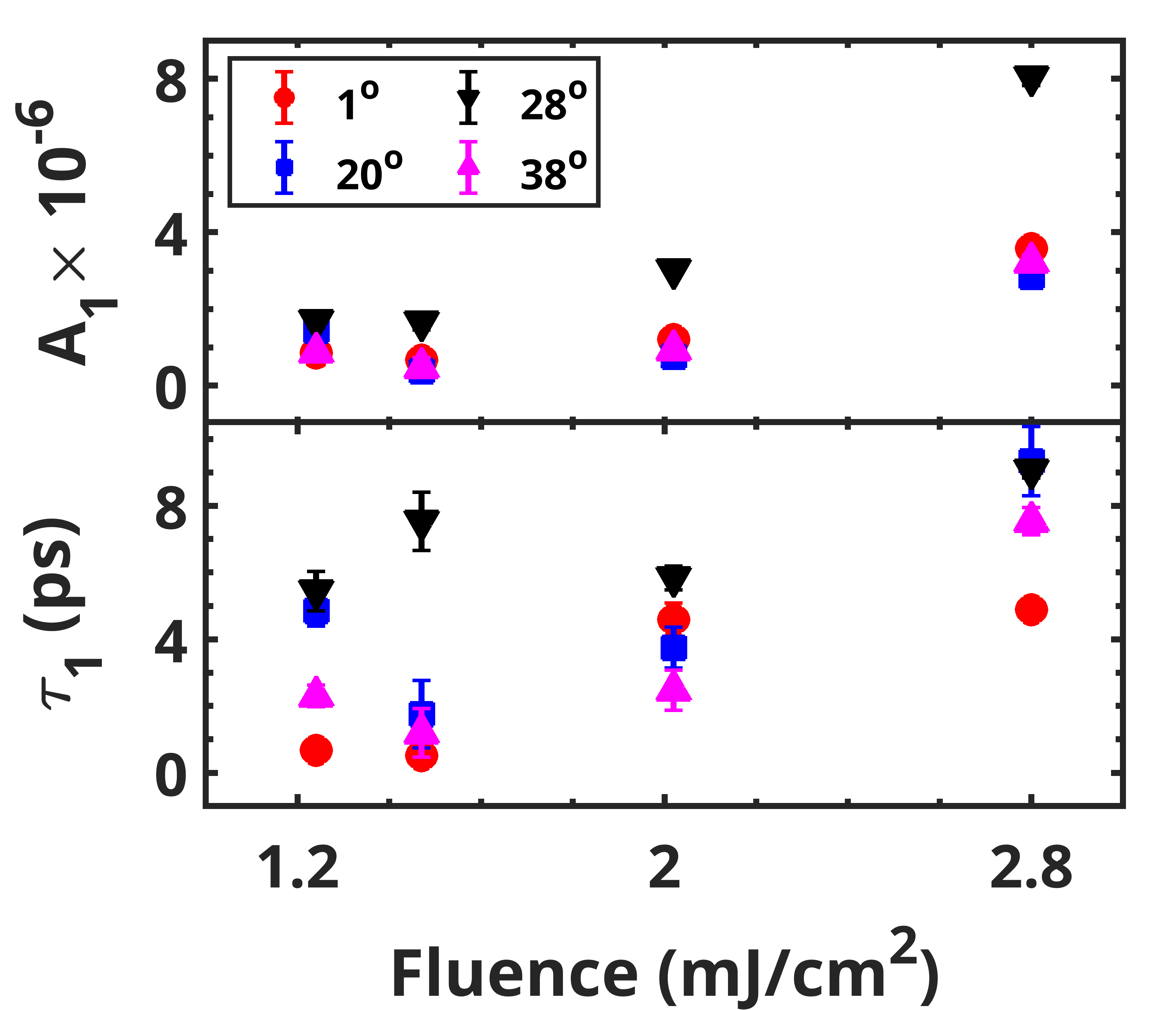}
    \renewcommand{\thefigure}{S\arabic{figure}}
    \caption[Fluence dependence of amplitude A$_1$ and relaxation time $\tau_1$ at different twist angles]{Fluence dependence of amplitude A$_1$ and relaxation time $\tau_1$ at different twist angles.}
    \label{Fig:6pS3}
\end{figure}